\documentclass{elsart3}
\usepackage{amssymb}
\usepackage{epsfig}

\begin{document}

\begin{frontmatter}

\title{Determination of an onset of superconducting diamagnetism by scaling of the normal-state magnetization}

\author{I. L. Landau, K. M. Mogare, B. Trusch, M. Wagner, J. Hulliger}

\address{Department of Chemistry and Biochemistry, University of Berne, Freiestrasse 3, CH-3012-Berne, Switzerland}

\begin{abstract}

We propose a simple scaling procedure for the normal-state magnetization $M_n$ data collected as functions of temperature $T$ in different magnetic fields $H$. As a result, the $M_n(T)$ curves collected in different fields collapse on to a single $M_{sc}(T)$ line. In this representation, the onset of superconducting diamagnetism manifests itself by a sharp divergence of the $M_{sc}(T)$ curves for different $H$ values. As will be demonstrated, this allows for a reliable determination of temperature $T_{onset}$, at which superconducting diamagnetism become observable.

\end{abstract}

\end{frontmatter}

\section{Introduction}

An onset of superconducting diamagnetism in a vicinity of the superconducting critical temperature $T_c$ attracts a lot of attention in high-$T_c$ superconductors \cite{1,2,3,4,5,6,7,8,9}. This attention is partly dictated by an obvious fundamental interest to the onset of superconductivity in such complex materials. It is also important from purely practical reasons because it provides the highest value of $T_c$ in non-uniform superconducting samples.

At the same time, in spite of apparent simplicity of the problem, experimental studies are rather complex. Because all high-$T_c$ superconductors have a considerable temperature dependence of the normal-state magnetization $M_n$, extrapolation of $M_n$ to a region of the superconducting transition is not obvious and sometimes questionable (see, for instance, Fig 1 of Ref. \cite{8} and Figs. 8 and 9 in Ref. \cite{9}). 

\begin{figure}[!t]
 \begin{center}
 \epsfxsize=0.95\columnwidth \epsfbox {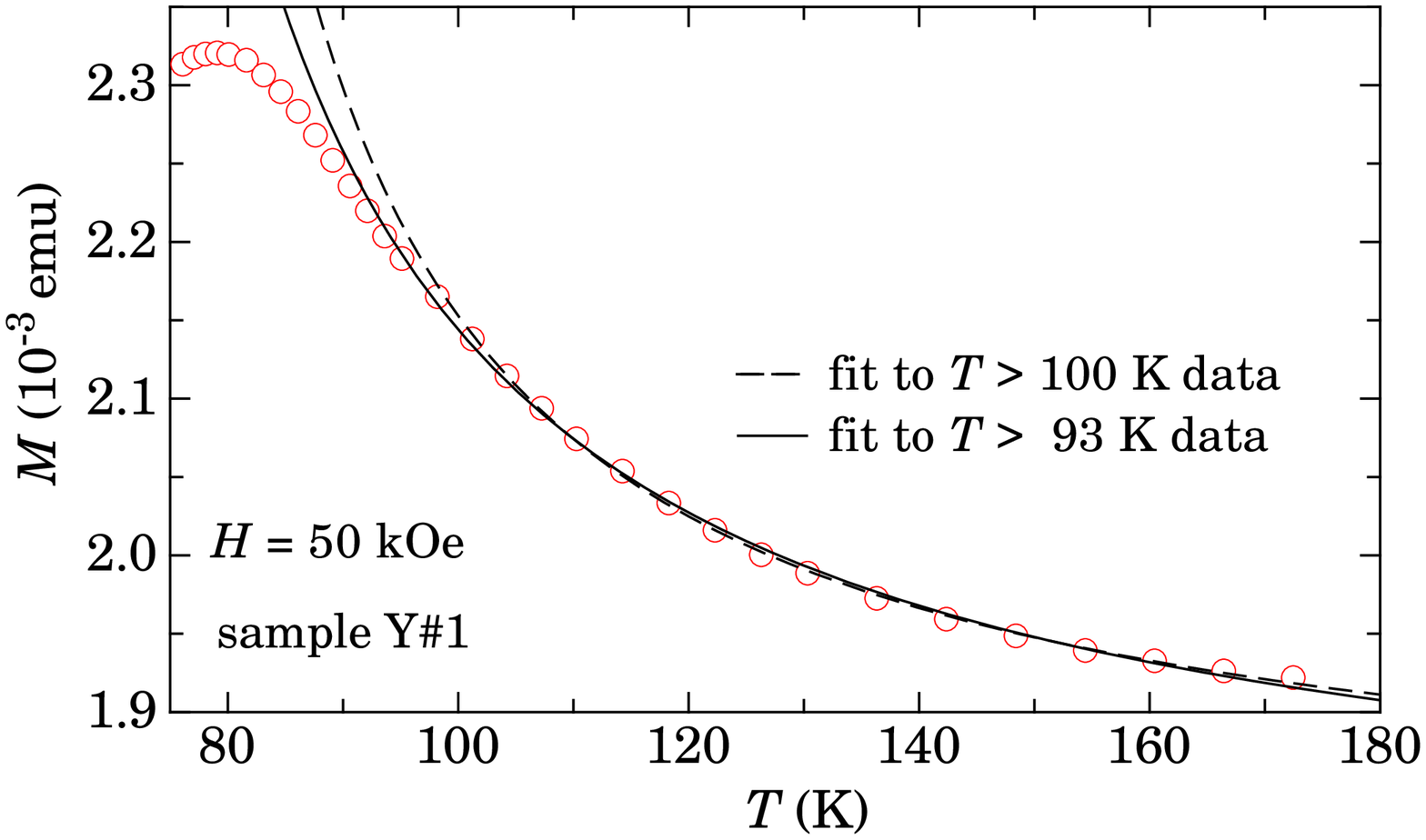}
  \caption{Magnetization $M(T)$ data for Y\#1 sample collected in a magnetic field $H = 50$ kOe. The solid and the dashed lines are fits of Eq. (1) for two sets of experimental data points. }
 \end{center}
\end{figure}
As an example, Fig. 1 shows magnetization data for an oxygen depleted sample of YBa$_2$Cu$_3$O$_{7-x}$. Detailed data for this sample will be presented and discussed below. We have tried to fit experimental  $M_n(T)$ data by the Curie-Weisse law plus some temperature independent constant $M_0$:
\begin{equation}
\chi = C/(T- \Theta) + M_0.
\end{equation}
$C$, $\Theta$ and $M_0$ were used as fit parameters. As may be seen in Fig. 1, the calculated curves do not fit experimental data points particularly well. Furthermore, the corresponding values of the onset temperature $T_{onset}$ noticeably depend on the data, which were chosen for approximation. In such and similar situations any conclusion about the onset of superconductivity is not reliable.
 
Here, we shall introduce a simple procedure to scale the normal-state magnetization data collected in different magnetic fields. Because the proposed procedure does not involve any specific assumption about sample properties, it is applicable to uniform and non-uniform samples, to single crystals and ceramics.

\section{Samples}

\begin{figure}[h]
 \begin{center}
 \epsfxsize=0.95\columnwidth \epsfbox {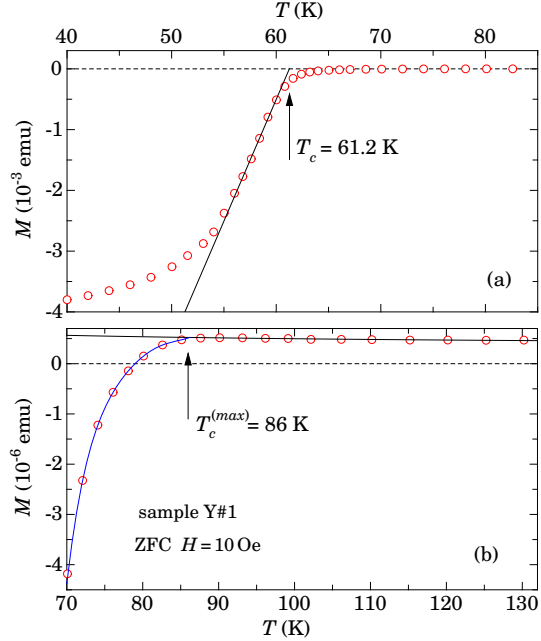}
  \caption{Magnetization $M(T)$ data for Y\#1 sample collected in a low magnetic field $H = 10$ Oe after zero-field-cooling (ZFC). (a) The main part of the transition. The solid line is a linear fit to the steepest part of the $M(T)$ curve. (b) A tale of the transition with $y$-scale expanded by three orders of magnitude. The solid line is a linear fit to $M(T)$ data at $T>90$ K. Definitions of $T_c$ and $T_c^{max}$ are shown in Figs. 2(a) and 2(b), respectively. }
 \end{center}
\end{figure}

In this section we present a brief description of samples, which were used to verify the proposed scaling procedures. Two kinds of high-$T_c$ materials were investigated: YBa$_2$Cu$_3$O$_{7-x}$ (YBCO) and Tl$_2$Ba$_2$Ca$_2$Cu$_3$O$_{10}$ (Tl-2223). All samples were ceramics and were not particularly uniform. Fig. 2 shows a superconducting transition for an oxygen depleted sample Y\#1. The mean-field superconducting temperature $T_c$ is defined as is shown in Fig. 2(a). Because this sample was kept on air for about a year, it contained inclusions with a higher level of doping and with correspondingly higher $T_c$ values.   This resulted in about 25 K long tale of the transition (see Fig. 2(b)). Definitions of the mean-field critical temperature $T_c$ and $T_c^{(max)}$ are shown in Fig.2. Main characteristics of samples are summarized in Table 1.

\begin{table}[h]
 \centering 
\caption{Main parameters of samples. The effective magnetic field $h_{f-p}$ characterizes a relative strength of  a ferromagnetic contribution to the sample magnetization (see Eq. (4) for the definition)}
\begin{tabular}{lcccccc}
\hline
\hline
 sample  & compound & $T_c$ & $T_c^{max} $ & $h_{f-p}$ & $T_{onset}$ (K)    \\
 \hline
 \hline
Y\#1  &  YBCO  & 61.2 K & 86 K & 0.14 kOe & $88.5\pm0.5$ \\
 & underdoped & & &\\
\hline
Y\#2 &  YBCO & 91.1 K & 92 K & 0.9 kOe & $92.3\pm0.2$ \\
\hline 
Y\#3 &  YBCO  & 91.2 K & 92 K & 1 kOe & $92.5\pm0.3$ \\
\hline
Tl\#1 &  Tl-2223  & 122 K & 130 K & 0 & $131\pm0.3$ \\
\hline
Tl\#2 &   Tl-2223  & 100 K & 125 K & 2.9 kOe & $125\pm 1$\\
\hline
\hline
\end{tabular}
\end{table}

\section{Scaling procedure}

We consider two modifications of the scaling procedure. A simpler version is applicable to samples with purely paramagnetic normal-state magnetization. If the normal-state magnetization includes also a ferromagnetic contribution, the procedure should be modified in oder to account for a nonlinearity of $M(H)$ curves. 
 
\subsection{Purely paramagnetic case}

Because paramagnets may be described by a magnetic field independent magnetic susceptibility $\chi_{p}$, the sample magnetization $M = \chi_{p}H$, i.e., $M$ is proportional to $H$ at any temperature. This means that $M_n(T)$ curves collected in different fields will collapse onto a single $M_{sc}(T)$ curve  if 
\begin{equation}
M_{sc}(H)=\frac{M(H,T)}{M(H,T_0)},
\end{equation}
where $T_0$ is any temperature well above $T_c$.

\begin{figure}[h]
 \begin{center}
 \epsfxsize=0.95\columnwidth \epsfbox {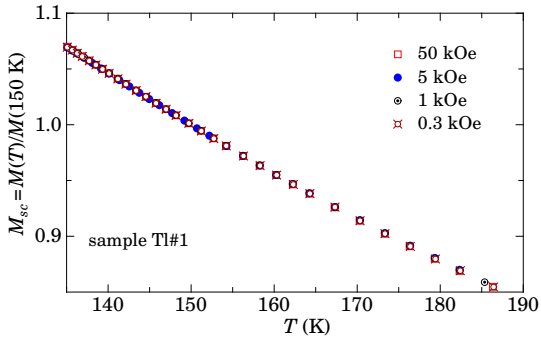}
  \caption{Magnetization data for Tl$\#$1 sample collected in four different fields plotted as $M_{sc}=M(T)/M(150$ K$)$ versus $T$}.
 \end{center}
\end{figure}
Fig. 3 illustrates results of such a scaling. Agreement between the data is practically perfect. While magnetic fields differ by more than two orders of magnitude, the differences between the $M_{sc}(T)$ curves do not exceed an experimental scatter.

\begin{figure}[h]
 \begin{center}
 \epsfxsize=0.95\columnwidth \epsfbox {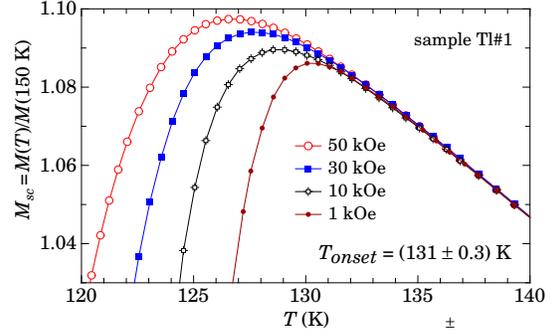}
  \caption{ The scaled magnetization data for the Tl$\#$1 sample in the transition region.  The solid lines are guides to the eye. }.
 \end{center}
\end{figure}
The scaled magnetization curves for lower temperatures are shown in Fig. 4. It may be seen that the onset temperature $T_{onset} \approx 131$ K. There is some experimental uncertainty, but there are no systematic errors, which may distort the conclusions. 

\subsection{Paramagnetism with a ferromagnetic contribution}

\begin{figure}[h]
 \begin{center}
 \epsfxsize=0.95\columnwidth \epsfbox {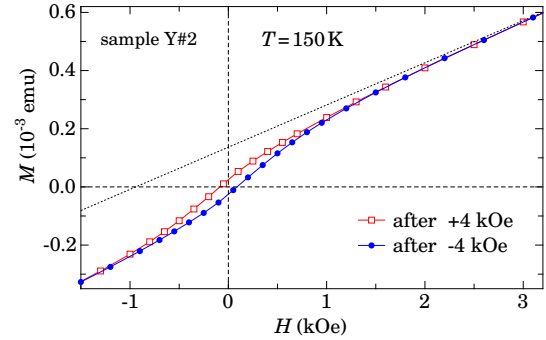}
  \caption{Magnetization data for the sample Y\#2 measured at $T=150$ K in fields $-4$ kOe $<H<4$ kOe. The solid lines are guides to the eye. The dotted line is a linear fit to data collected in fields $5$ kOe $<H<50$ kOe.}.
 \end{center}
\end{figure}
Quite often, the normal-state magnetization of high-$T_c$ superconductors cannot  be described as purely paramagnetic. A typical example is shown in Fig. 5. There is a quite symmetrical hysteresis lope with non-zero magnetizations at $H=0$, which is an unambiguous signature of ferromagnetism.  

\begin{figure}[h]
 \begin{center}
 \epsfxsize=0.95\columnwidth \epsfbox {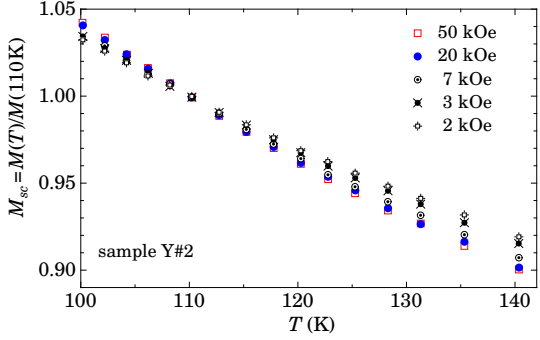}
  \caption{$M(T)/M(110$ K$)$ measured in different fields as functions of temperature.}.
 \end{center}
\end{figure}
A simple scaling procedure, which was introduced in the previous subsection does not work, as may clearly be see in Fig. 6. Because, the sample magnetization $M(H)=\chi H$ cannot be described with $\chi$ independent of field, this failure is expected.

\begin{figure}[h]
 \begin{center}
 \epsfxsize=0.95\columnwidth \epsfbox {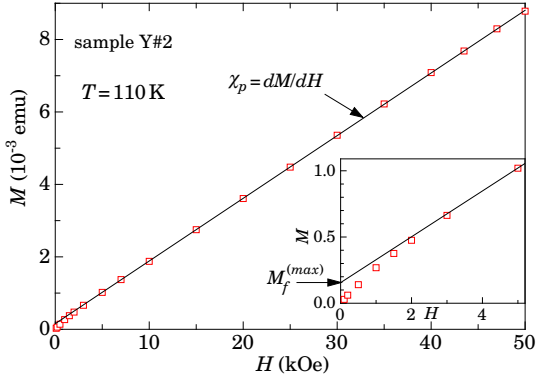}
  \caption{$M$ at $T=110$ K versus field. The solid line is a linear fit to the data for $H\ge5$ kOe. The inset shows the low-field part of the curve on expanded scales. Arrows illustrate definitions of $\chi_p$ and $M_f^{(max)}$ }.
 \end{center}
\end{figure}
If there are several contributions to the sample magnetization, $M$ may be written as a sum. In our case,
\begin{equation}
M = \chi_{p}H + M_{f},
\end{equation}
where indexes $p$ and $f$ are related to paramagnetic and ferromagnetic contributions, respectively. Fig. 7 shows $M(H)$ data for the same sample Y\#2 measured in more details. As may be seen, for $H \ge 5$ kOe, $M$ is a linear function of $H$. This allows to conclude that all ferromagnetic inclusions are already saturated and in $H\ge 5$ kOe $M_{f}$ is equal to its maximum value $M_{f}^{(max)}$. All calculations for this sample that we present below were made for $T_0=110$ K. $\chi_{p}(T_0)=dM/dH \approx 1.73\cdot 10^{-4}$  emu/kOe and $M_f^{(max)}(T_0) \approx 1.56\cdot 10^{-4}$.  In lower fields, $M_f$ depends not just on field, but also on the magnetic history (see Fig. 4). However, $M_f(H)$ can easily be calculated by employing of Eq. (3). 

In order to characterize a relative strength of a ferromagnetic contribution, we introduce
\begin{equation}
h_{f-p} = \frac{M_f^{(max)}}{\chi_p}.
\end{equation}
The value of this effective field is a convenient way to characterize ferromagnetic contributions to $M$. The values of $h_{f-p}$ for the investigated samples are presented in Table 1. 

As the next step, we assume that $M_f(H)$ is temperature independent for all $H$ values. This is the only realistic way to scale the normal-state magnetization data without making rather specific assumptions about sample properties, which may distort the final result. Although this cannot be exact, we shall show that in some limited temperature range above $T_c$ this assumption works  sufficiently well to ensure satisfactory scaling of experimental data. Thus, using Eq (3) and the value of $\chi_p(T_0)$, we calculate $M_f(H,T_0)=M-\chi_pH$. Then, experimental $M(T)$ data for different fields are scaled according to
\begin{equation}
M_{sc}(T) =\frac{M(H,T) - M_f(H,T_0)}{\chi_p(T_0)H} 
\end{equation}

\begin{figure}[h]
 \begin{center}
 \epsfxsize=0.95\columnwidth \epsfbox {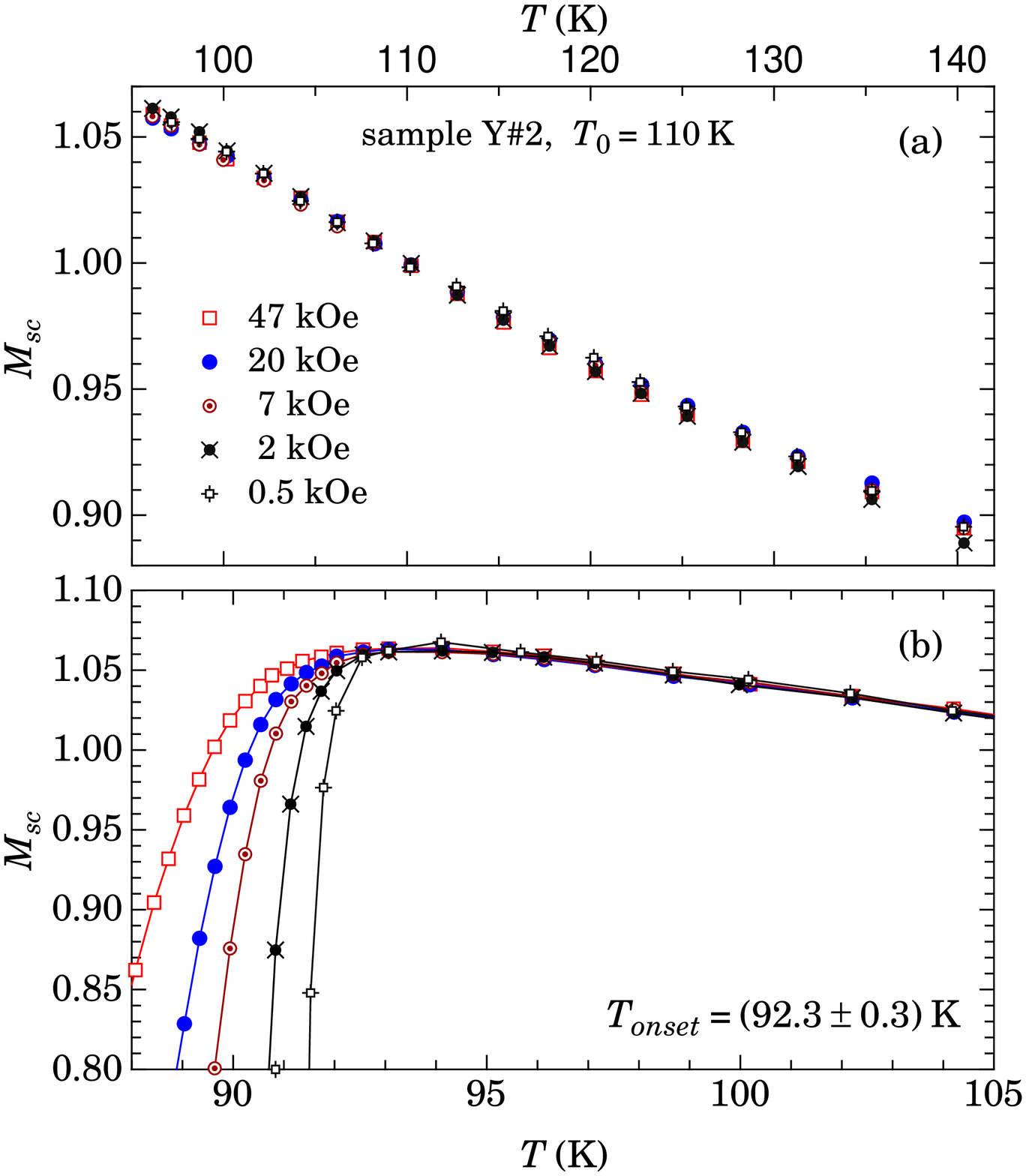}
  \caption{Magnetization data for the sample Y\#2 scaled according to Eq. (5) versus temperature. (a) For $T>T_c$. (b) In the transition region. The solid lines are guides to the eye.}.
 \end{center}
\end{figure}
The results of such scaling with $T_0=110$ K are presented in Fig. 8(a). As may be seen, $M_{sc}(T)$ curves calculated from experimental $M(T)$ data measured in different fields perfectly match each other in a rather extended temperature range $T>T_c$. The onset of diamagnetism is shown in Fig. 7(b). It can easily be established that $92\texttt{ K}<T_{onset}<92.7$ K. 

\begin{figure}[h]
 \begin{center}
 \epsfxsize=0.95\columnwidth \epsfbox {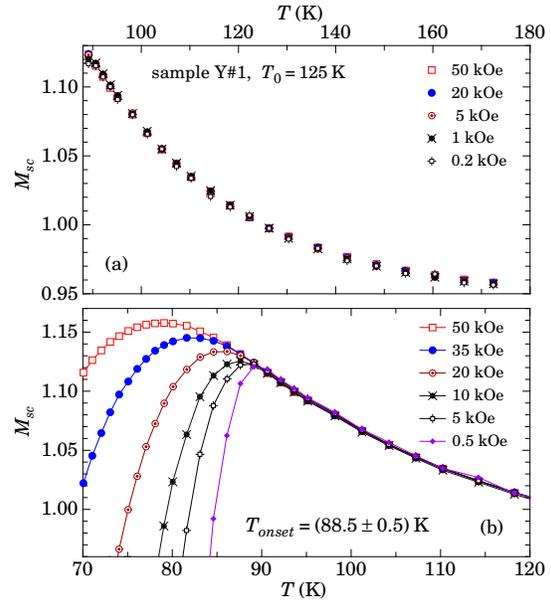}
  \caption{$M_{sc}$ for the sample Y\#1 versus temperature. (a) For $T>T_c$. (b) In the transition region. The solid lines are guides to the eye}.
 \end{center}
\end{figure}
Similar results for the oxygen deficient sample Y\#1 are shown in Fig. 9. As may be seen, $T_{onset} = (88.5\pm0.5)$ K.  For this sample,  $h_{f-p} = 0.14$ kOe, i.e., a relative ferromagnetic contribution is  approximately 6 times smaller than that for the sample Y\#1.  

\begin{figure}[h]
 \begin{center}
 \epsfxsize=0.95\columnwidth \epsfbox {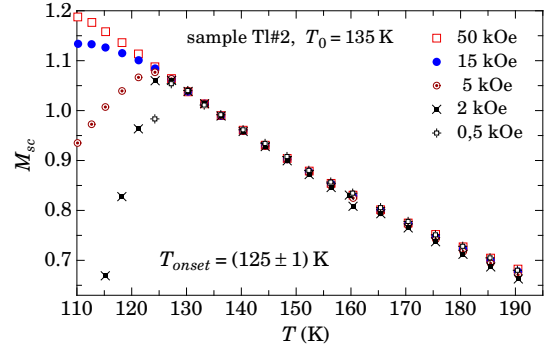}
  \caption{$M_{sc}$ versus $T$ for the sample Tl\#2. }.
 \end{center}
\end{figure}
Generally, a small ferromagnetic contribution favors good quality of scaling. However, the most important factor is not a low value of $h_{f=p}$, but rather temperature independence of $M_f(H)$, which is the basis of our scaling approach. For instance, for the sample Tl\#2, $h_{f-p} = 3$ kOe, i.e., considerably higher than for all other samples presented here. Nevertheless, as may be seen in Fig. 10, $M_{sc}(T)$ curves calculated from data measured in different fields perfectly match each other in a rather extended ranges of temperatures and fields.

\begin{figure}[h]
 \begin{center}
 \epsfxsize=0.95\columnwidth \epsfbox {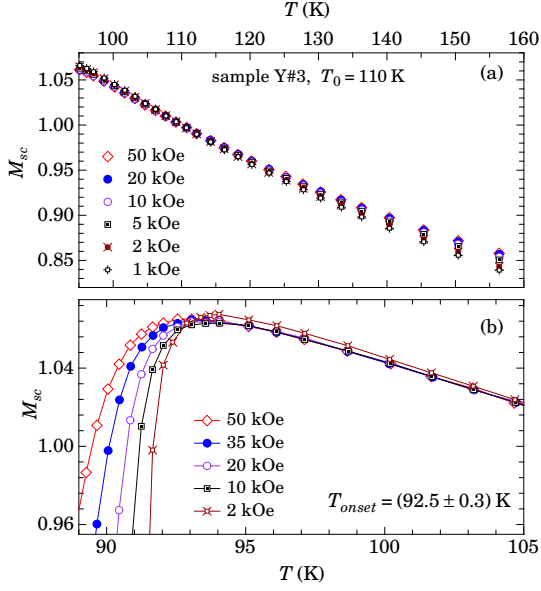}
  \caption{$M_{sc}$ versus $T$ for the sample Y\#2. (a) For $T>T_c$. (b) In the transition region. The solid lines are guides to the eye. }.
 \end{center}
\end{figure}
There were, however, some cases, in which the quality of scaling was not as good as presented above. The results for such a sample are shown in Fig. 11. While formal characteristics of this sample, including the value of $h_{f-p}$ are quite similar to that of the sample Y\#2 (see Table 1), quality of scaling for the sample Y\#3 is not the same good (compare Figs. 8(a) and 11(a)). In the case of the sample Y\#3, the data for $H \le 5$ kOe deviate downwards from the master curve at higher temperatures (see Fig. 11(a)). Similar deviations upwards are visible at lower temperatures (see Fig. 11(b)).

\begin{figure}[h]
 \begin{center}
 \epsfxsize=0.95\columnwidth \epsfbox {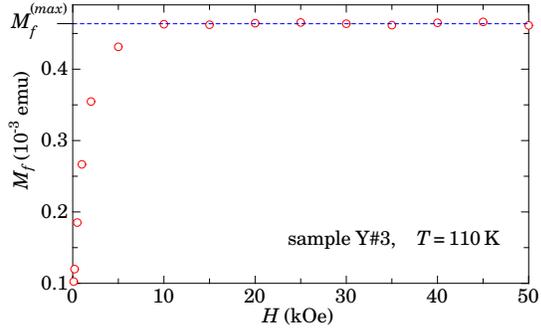}
  \caption{$M_f= M - \chi_pH$ at $T=110$ K as a function of $H$. The dashed line corresponds to $M_f^{max}=1.64\cdot 10^{-4}$ emu. }.
 \end{center}
\end{figure}
The obvious reason is a noticeable dependence of $M_f(H)$ on temperature. We remind that the temperature independence of $M_f(H)$ is the basis of this scaling procedure.  At the same time, the data for $H\ge10$ kOe can be scaled quite well (see Fig. 11). This is evidence that, while $M_f$ depends on temperature in lower fields, the saturated value of the ferromagnetic contribution to the sample magnetization $M_f^{(max)}$ is practically temperature independent. Indeed, as may be seen in Fig. 12, $M_f$, calculated according to Eq. (3), is independent of field down to $H=10$ kOe, while the data point for $H=5$ kOe is already substantially below $M_f^{(max)}$.

\subsection{Analysis of $M_{sc}(H,T)$ data below $T_{onset}$}

Because the divergence between the $M_{sc}(T)$ curves manifests the onset of superconducting diamagnetism, the difference 
\begin{equation}
\Delta M_{sc} = M_{sc}(H) - M_{sc}(50\texttt{ kOe})
\end{equation}
may serve as some kind of its quantitative characteristic. 

\begin{figure}[h]
 \begin{center}
 \epsfxsize=0.95\columnwidth \epsfbox {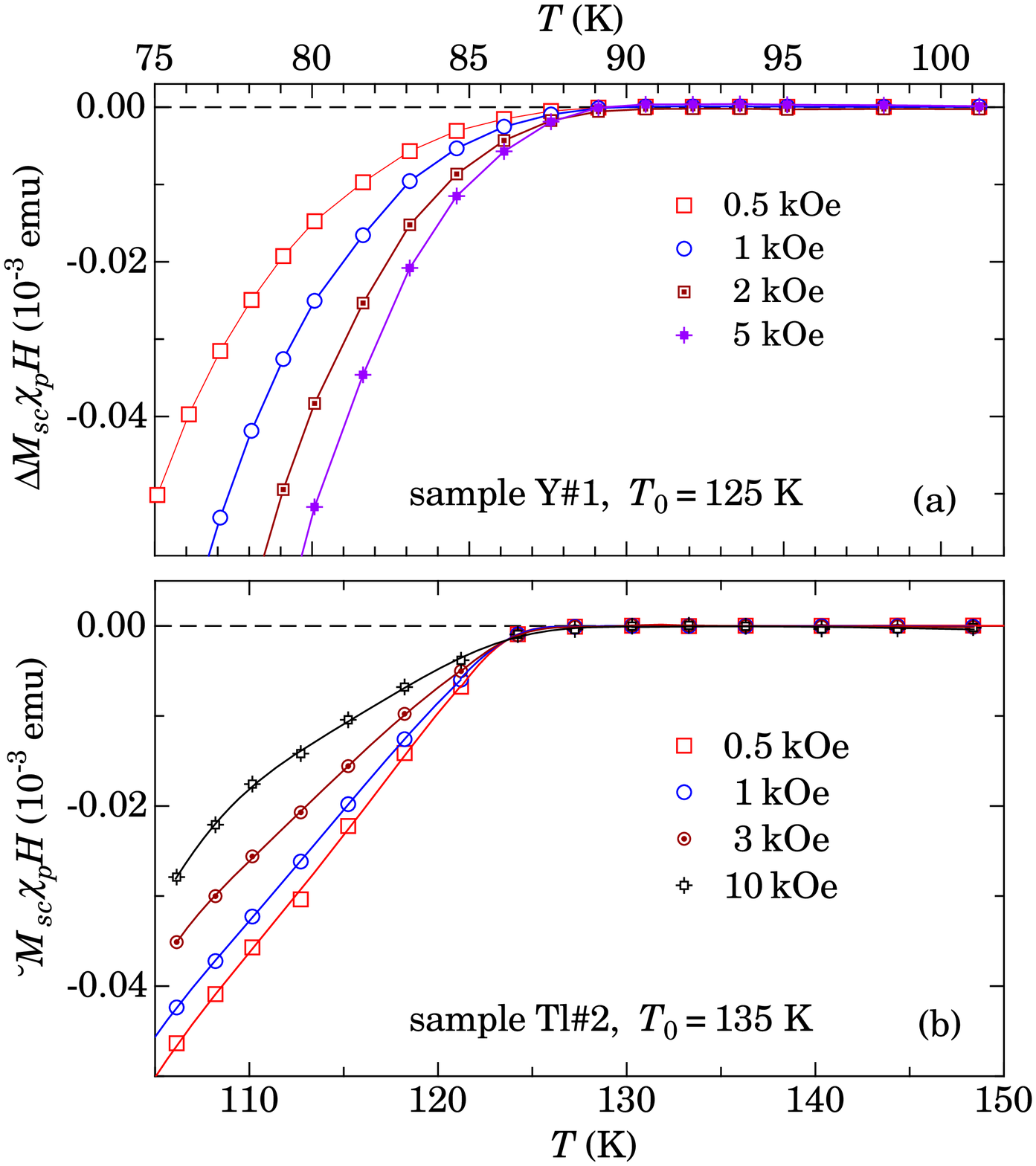}
  \caption{ $\Delta M_{sc}\chi_pH$ versus $T$. (a) For the sample Y\#1. (b) For  Tl\#2 The solid lines are guides to the eye. }. \end{center}
\end{figure}
In order to compare $\Delta M_{sc}$ results for different values of $H$, it is convenient  to use $\Delta M_{sc}\chi_pH$ (see Eq. (5)). Fig. 13 shows the $\Delta M_{sc}\chi_pH$ curves for the sample Y\#1 and Tl\#2. The values of $T_{onset}$ resulting from this representation of the results are the same as may be determined from Figs. 10 and 11. In this case, the main source of uncertainty in $T_{onset}$ is the distance between the neighboring data points. We also note a rather drastic difference in $\Delta M_{sc}(H)\chi_pH$ dependencies below $T_{onset}$ for these two samples. For the sample Y\#1, $\Delta M_{sc}(H)\chi_pH$ is an increasing function of $H$ (Fig. 13(a)). Contrary to that, $\Delta M_{sc}(H)\chi_pH$ decreases with increasing $H$. Quite likely, this is one of manifestations of general differences between Y- and Tl-based cuprates.

\section{Conclusions}

The proposed approach allows to scale the normal-state magnetization $M_n(T)$ data in a way that the $M_{sc}(T)$ curves calculated from experimental $M_n(T)$ data collected in different magnetic fields collapse onto the same master curve. Because a diamagnetic contribution to $M$ due to superconductivity depends on magnetic field in a way, which is quite different from that for the normal-state magnetization, the onset of superconductivity leads to a rather pronounced divergence between the $M_{sc}(T)$ curves corresponding to different magnetic fields, as it may clearly be seen in Figs. 4, 8, 9, 10, 11(b) and 13. This allows for an unambiguous determination of $T_{onset}$ corresponding to the onset of superconducting diamagnetism. Accuracy of $T_{onset}$ is determined by an experimental scatter of $M(T)$ data and by distances between neighboring $M(T)$ data points. We remind that the proposed approach relies only on rather general properties of the normal-state magnetization and does not include any specific assumptions, which cannot be independently verified.  We also note that the possibility to check the quality of scaling at temperatures well above $T_{oncet}$ serves as some consistency check. If the scaling procedure does not work satisfactory, the resulting data should not be used for important conclusions.

There are two main reasons to have quite considerable values of $(T_c - T_{onset})$. (i) Thermal fluctuations, which are expected to be especially strong in high-$T_c$ superconductors and (ii) inclusions of small quantities of phases with higher values of $T_c$. Here, we mainly consider technical aspects of scaling of the normal-state magnetization data and discussion of a possible nature of the transition at $T=T_{onset}$ in different samples is beyond the scope of this paper.

\ack 

This work was supported by the Swiss NCCR MaNEP II under project 4, novel materials.

\end{document}